\documentclass[aip,reprint,graphicx]{revtex4-1}
\usepackage{graphicx,color}
\usepackage{amsmath, amssymb}
\usepackage{longtable}
\usepackage{multirow}
\usepackage{natbib}
\usepackage{bm}
\allowdisplaybreaks[1]

\begin{document}
\title{Dynamical Jahn-Teller effect in the first excited C$_{60}^-$} 
\author{Zhishuo Huang}
\email[]{zhishuohuang@gmail.com}
\affiliation{Theory of Nanomaterials Group, KU Leuven, Celestijnenlaan 200F, B-3001 Leuven, Belgium}

\author{Dan Liu}
\email[]{iamdliu@nwpu.edu.cn}
\affiliation{Shaanxi Institute of Flexible Electronics, Northwestern Polytechnical University, 127 West Youyi Road, Xi'an, 710072, Shaanxi, China.}
\affiliation{Theory of Nanomaterials Group, KU Leuven, Celestijnenlaan 200F, B-3001 Leuven, Belgium}

\date{\today}

\begin{abstract}
The Jahn-Teller effect of C$_{60}$ anions in the first electronically excited states was theoretically investigated. 
The orbital vibronic coupling parameters for the $t_{1g}$ next lowest unoccupied molecular orbitals were derived from the Kohn-Sham orbital levels with hybrid B3LYP functional by using the frozen phonon approach. 
With the use of these coupling parameters, the vibronic states of the first excited C$_{60}^-$ were derived by exactly diagonalizing the dynamical Jahn-Teller Hamiltonian.
The dynamical Jahn-Teller stabilization energy of the first excited C$_{60}^-$ is stronger than that of the ground electronic states. 
\end{abstract}


\maketitle 

\section{Introduction}
Highly symmetric C$_{60}$ \cite{Kroto} exhibits complex Jahn-Teller (JT) dynamics characterized by orbital-vibration entanglement in various charged and excited states \cite{Chancey1997, Bersuker2006, Dunn2015}. 
Negatively charged C$_{60}$ has been one of the most investigated because it forms various molecular crystals \cite{Gunnarsson2004, Capone, Alloul, Kamaras, Takabayashi, Nomura2016, Otsuka}. 
Since the molecular nature strongly remains in these materials, for the thorough understanding of the JT effect of C$_{60}$ anions is crucial. 
Though JT effect, including dynamic JT one, of C$_{60}$ anions has been intensively investigated \cite{Auerbach1994, Manini1994, Yu1994, Dunn1995, Gunnarsson1995, OBrien1996, Tosatti1996, Yu1997, Manini1998, Sookhun2003, Dunn2005, Tomita2005, Hands2008, Frederiksen2008, Iwahara2010, Dunn2012, Klupp2012, Stochkel2013, Ponzellini, Kundu2015, Iwahara2018, Liu2018a, Liu2018b, Matsuda2018}, it is only last years that the actual situation in the ground electronic states of C$_{60}^{n-}$ molecule $(n = 1-5)$ has been established with accurate coupling parameters \cite{Liu2018a, Liu2018b}.

The JT effect is also considered to be important in the excited states of C$_{60}$ anions.
For example, the JT effect in the first excited C$_{60}$ anion where the $t_{1g}$ next lowest unoccupied molecular orbitals (NLUMOs) is populated is of fundamental importance to interpret absorption spectra of isolated C$_{60}^-$ \cite{Kato1991, Kato1993, Kodama1994, Kondo1995, Kwon2001, Kwon2002, Tomita2005, Stochkel2013, Watariguchi2016}, electron transfer process of fullerene \cite{ET1, ET2}, and excitation spectra of alkali-doped fullerides \cite{Knupfer1997, Chibotaru1999, Chibotaru2000}. 
The importance is also suggested \cite{Nava2018} in recently reported light induced superconductivity of alkali-doped fullerides \cite{Mitrano2016, Cantaluppi2017}. 
Moreover, the JT effect involving the NLUMO must be significant in highly alkali doped \cite{Knupfer1997} and alkali-earth/rare-earth doped fullerides \cite{Chen1999, Margadonna2000, Iwasa2003, Li2003, He2005, Akada2006, Heguri2010}. 

So far, the dynamic JT effect in negatively charged C$_{60}$ in the ground electronic configuration where only the $t_{1u}$ LUMOs has been mainly investigated. 
Recently, bound excited states of C$_{60}^-$ have been theoretically investigated \cite{EX1,EX2,EX3,EX4,EX5}, and the stability of the first excited ${}^2T_{1g}$ electronic states of C$_{60}^-$ has been confirmed. 
Nevertheless, the JT effect in the excited C$_{60}$ has not been theoretically investigated, and the actual situation in C$_{60}$ anions remains unclear.

In this work, we address the JT effect of first excited C$_{60}^-$ anion of $t_{1g}^1$ configuration.
The vibronic coupling parameters are derived from the $t_{1g}$ orbital energy levels calculated by density functional theory (DFT) calculations with hybrid B3LYP exchange-correlation functional. 
Using these coupling parameters, the vibronic states are obtained by numerically diagonalizing the dynamical JT Hamiltonian matrix.
Compared with the case of the ground electronic state of C$_{60}^-$, $t_{1u}^1$, the stabilization in the present case is found to be stronger by about 20 \%.

\section{Jahn-Teller Effect}
\subsection{Model Hamiltonian}\label{Sec:H}
The $t_{1g}$ next LUMO of neutral C$_{60}$ with $I_h$ symmetry is triply degenerate and separated from the other orbital levels \cite{Chancey1997}.
According to the selection rule, the $t_{1g}$ orbitals couple to totally symmetric $a_g$ and five-fold degenerate $h_g$ normal modes as in the case of $t_{1u}$ orbitals \cite{Jahn1937}:
\begin{eqnarray}
 [t_{1g} \otimes t_{1g}] = a_g \oplus h_g.
\label{Eq:selection}
\end{eqnarray}
Therefore, the linear vibronic Hamiltonian of C$_{60}^-$ in the first excited $t_{1g}^1$ electronic configuration is given as in the case of $t_{1u}^1$ \cite{OBrien1969, Auerbach1994, OBrien1996, Chancey1997}:
\begin{eqnarray}
 H &=& H_a + H_h,
\label{Eq:H}
\\
 H_a &=&
  \frac{1}{2}
  \left(
  p_a^2 + \omega_a^2 q_{a}^2
 \right) +  V_a q_{a},
\label{Eq:Ha}
\\
 H_h &=&
  \sum_{\gamma = \theta, \epsilon, \xi, \eta, \zeta} \frac{1}{2}\left(p_{h\gamma}^2 + \omega_h^2 q_{h \gamma}^2\right)
\nonumber\\
 &&+
 V_h
 \begin{pmatrix}
  \frac{1}{2} q_{h\theta} - \frac{\sqrt{3}}{2} q_{h\epsilon} & \frac{\sqrt{3}}{2} q_{h\zeta} & \frac{\sqrt{3}}{2} q_{h\eta} \\
  \frac{\sqrt{3}}{2} q_{h\zeta} & \frac{1}{2} q_{h\theta} + \frac{\sqrt{3}}{2} q_{h\epsilon} & \frac{\sqrt{3}}{2} q_{h\xi} \\
  \frac{\sqrt{3}}{2} q_{h\eta}  & \frac{\sqrt{3}}{2} q_{h\xi} & -q_{h\theta} \\
 \end{pmatrix}.
\label{Eq:Hh}
\end{eqnarray}
We take the equilibrium structure of C$_{60}$ as the reference structure.
Here, $q_{\Gamma\gamma}$ and $p_{\Gamma\gamma}$ ($\gamma = \theta, \epsilon, \xi, \eta, \zeta$ for $\Gamma = h$) are mass-weighted normal coordinates \cite{Inui1990} and conjugate momenta, respectively, $\omega_\Gamma$ is frequency, and $V_\Gamma$ the vibronic coupling parameters. The basis of the marix is in the order of $|T_{1g}x\rangle$, $|T_{1g}y\rangle$, $|T_{1g}z\rangle$. 
The representation for the normal coordinates and conjugate momenta possess the symmetry of real $d$-type [$(2z^2-x^2-y^2)/\sqrt{6}$, $(x^2-y^2)/\sqrt{2}$, $\sqrt{2}yz$, $\sqrt{2}zx$, $\sqrt{2}xy$], as they are consistent with the original and most used representation \cite{OBrien1969, Auerbach1994, Manini1994, OBrien1996, Chancey1997}.
The bases are different from those ($Q$) of some previous work \cite{Dunn1995}: 
\begin{eqnarray}
 q_\theta =\sqrt{\frac{3}{8}} Q_\theta + \sqrt{\frac{5}{8}} Q_\epsilon, \quad
 q_\epsilon=\sqrt{\frac{3}{8}} Q_\theta - \sqrt{\frac{5}{8}} Q_\epsilon.
\end{eqnarray}
In the above equation, the indices $g$ or $u$ indicating the parity and the indices $\mu$ distinguishing the frequencies are omitted for simplicity. 
They are added when necessary. 

\subsection{Adiabatic potential energy surface}
The model Hamiltonians, and hence the formulae, for the ground electronic configuration and the first excited configuration have the same structure. 
The depth of the adiabatic potential energy surface (APES) with respect to the reference structure is given by \cite{OBrien1969}
\begin{eqnarray}
 U_\text{min} &=& -E_a - E_\text{JT}
\nonumber\\
 &=& -\frac{V_a^2}{2\omega_a^2} - \frac{V_h^2}{2\omega_h^2},
\label{Eq:Umin}
\end{eqnarray}
with
\begin{equation}
 q_{a,0} = -\frac{V_a}{\omega_{a}^{2}},
 \quad \left|\bm{q}_{h,0} \right|=\frac{V_h}{\omega_{h}^{2}},
\end{equation}
where $E_a$ and $E_\text{JT}$ are the first and the second terms in the last expression in Eq. (\ref{Eq:Umin}), respectively, and $\bm{q}_{h}$ is the list of $q_{h\gamma}$.
The APES has two-dimensional continuous trough \cite{OBrien1969}, suggesting the presence of SO(3) symmetry \cite{OBrien1971, Pooler1980}.

\subsection{Vibronic states}
\label{Sec:vibronic}
As in the case of the JT problem for the $t_{1u}^1$ configurations \cite{OBrien1971, Romestain1971, Pooler1980}, the vibronic angular momenta $\hat{\bm{J}}$ exist in the case of $t_{1g}^1$ \cite{Chancey1997}:
\begin{eqnarray}
 [\hat{H}_h, \hat{\bm{J}}^2] = [\hat{H}_h, \hat{J}_z] = 0.
\label{Eq:symmetry}
\end{eqnarray}
Therefore, the eignestates of $\hat{H}$ (vibronic states) are expressed by $J$, $M_J$, and principal quantum number $\alpha$,
\begin{eqnarray}
 \hat{H}_h|\alpha JM_J\rangle &=& E_{\alpha J} |\alpha JM_J\rangle.
\label{Eq:vibronicproblem}
\end{eqnarray}
The analytical treatments of the vibronic states in the strong limit of vibronic coupling \cite{OBrien1969, OBrien1971, Auerbach1994, OBrien1996, Iwahara2018} and weak coupling limit \cite{Manini1994} have been discussed much.
Nevertheless, for the quantitative description of C$_{60}$ ions, only numerical approach can provide accurate description.

For numerical calculations, it is convenient to expand the vibornic states as
\begin{eqnarray}
 |\alpha J M_J \rangle &=& \sum_\gamma \sum_{\bm{n}_h} |T_{1g} \gamma\rangle \otimes |\bm{n} \rangle C_{\gamma \bm{n}; \alpha J M_J}.
\end{eqnarray}
Here, $\bm{n}_h = (n_{h\theta}, n_{h\epsilon}, n_{h\xi}, n_{h\eta}, n_{h\zeta})$ is the set of vibrational quantum numbers of the Harmonic oscillation part of Eq. (\ref{Eq:Hh}).
Such an expansion using the direct products of the electronic states and the eigenstates of harmonic oscillator has been proposed long time ago \cite{Longuet-Higgins1958} and has been routinely used as a reliable approach to quantitatively study the dynamical JT systems including fullerene anions \cite{OBrien1971, Auerbach1994, Gunnarsson1995, OBrien1996, Iwahara2010, Iwahara2013, Ponzellini, Liu2018a}.

In the present calculations, the vibrational basis is truncated as
\begin{eqnarray}
 0 \le n_{h(\mu)\gamma}, \quad \sum_{\mu\gamma} n_{h(\mu)\gamma} \le 7,
\end{eqnarray}
because the dimension of the Hamiltonian matrix rapidly increases.
To take account of the eight sets of $h_g$ modes in real C$_{60}$, $\mu$ is shown.
For the diagonalization of the vibronic Hamiltonian (\ref{Eq:Hh}), Lanczos algorithm was employed \cite{Pooler1984}.

\subsection{Orbital vibronic coupling parameters}
\begin{figure}[bt]
\includegraphics[width=8cm]{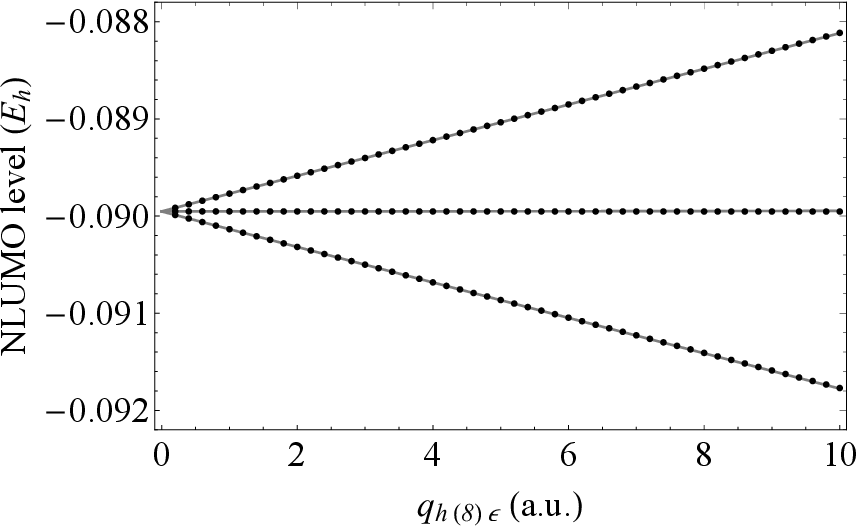}
\caption{
The JT splitting of the NLUMO levels with respect to $q_{h(8)\epsilon}$ deformation (in atomic unit).
The black points and gray lines indicate the DFT values and model energy, respectively.
}
\label{Fig:V}
\end{figure}

\begin{table*}
\caption{
The frequencies $\omega_\Gamma$ (cm$^{-1}$), vibronic coupling parameters $V_\Gamma$ (10$^{-4}$ a.u.), dimensionless vibronic coupling parameters $g_\Gamma = V_\Gamma/\sqrt{\hslash\omega_\Gamma^3}$, and stabilization energies $E_\Gamma$ (meV) for the $a_g$ and $h_g$ modes.
The data for LUMO are taken from Ref. \cite{Liu2018b} and the frequencies are taken from Ref. \cite{Bethune1991}.}
\label{Table:V}
\begin{tabular*}{\textwidth}{@{\extracolsep{\fill}}lllllllll}
\hline
     $J$    &       &                   & \multicolumn{3}{c}{NLUMO}& \multicolumn{3}{c}{LUMO}\\
     \hline
$\Gamma$ & $\mu$ & $\omega_{\Gamma}$ & $V_\Gamma$ & $g_\Gamma$ & $E_\Gamma$ & $V_\Gamma$ & $g_\Gamma$ & $E_\Gamma$\\
\hline
$a_g$ & 1 &  496 & $-0.449$ & $-0.418$ &  5.38 &$-0.264$  &$-0.245$  &  1.849 \\
      & 2 & 1470 & $-2.480$ & $-0.452$ & 18.66 &$-2.380$  &$-0.422$  & 16.543 \\
      \hline
$h_g$ & 1 &  273 & $-0.406$ & $-0.926$ & 14.50 & 0.192 &0.455  &  3.415 \\
      & 2 &  437 & $-0.476$ & $-0.536$ &  7.78 & 0.450 &0.503  &  6.886 \\
      & 3 &  710 & $-1.061$ & $-0.577$ & 14.64 & 0.754 &0.396  &  7.069 \\
      & 4 &  774 & $-0.594$ & $-0.284$ &  3.86 & 0.554 &0.259  &  3.256 \\
      & 5 & 1099 & $-0.498$ & $-0.141$ &  1.35 & 0.766 &0.209  &  3.038 \\
      & 6 & 1250 & $-1.664$ & $-0.387$ & 11.61 & 0.578 &0.132  &  1.360 \\
      & 7 & 1428 &   0.125  &   0.024  &  0.05 & 2.099 &0.394  & 13.867 \\
      & 8 & 1575 & $-2.113$ & $-0.348$ & 11.79 & 2.043 &0.326  & 10.592 \\
      \hline
\end{tabular*}
\end{table*}

The orbital vibronic coupling parameters are defined by the gradients of the $t_{1g}$ NLUMO level:
\begin{eqnarray}
 v_a = \left.\frac{\partial \epsilon_{t_{1g}z}}{\partial q_a}\right|_{\bm{q} = \bm{0}},
 \quad v_h = -\left.\frac{\partial \epsilon_{t_{1g}z}}{\partial q_{h\theta}}\right|_{\bm{q} = \bm{0}},
\end{eqnarray}
where $\bm{q}$ is the set of all normal coordinates.
In the present case, the vibronic coupling parameters $V_\Gamma$ correspond to the orbital vibronic coupling parameters $v_\Gamma$:
\begin{eqnarray}
 V_\Gamma = v_\Gamma,
\end{eqnarray}
in a good approximation because of the very small mixing of the orbitals under JT deformation.

The vibronic coupling parameters are derived by fitting the model potential to the gradients of NLUMO levels of neutral C$_{60}$ calculated in Ref. \onlinecite{Liu2018b}. 
The derivations were done using the DFT data with hybrid B3LYP functional because in the studies of C$_{60}$ anions, this functional has been shown to give the coupling parameters close to those derived from experimental data \cite{Iwahara2010} [In the study, high-resolution photoelectron spectra \cite{Wang2005} was used]. 
The coupling parameters derived from the B3LYP data are in good agreement with those from the gradients of the GW quasiparticle levels \cite{Faber2011}.
Furthermore, with the use of these parameters, the spin gap of C$_{60}^{3-}$ was well reproduced \cite{Liu2018a}.
The vibronic coupling parameters of C$_{60}^-$ have been derived using local density approximation \cite{Manini2001}, generalized gradient approximation \cite{Frederiksen2008}, and also by post Hartree-Fock calculations \cite{Iwahara2012}. 
Nevertheless, the former two methods underestimate and the latter one overestimates the vibronic coupling parameters. 
Therefore, we expect that the vibronic coupling parameters derived from the gradient of the NLUMO levels are accurate enough to reveal the low-energy states of excited C$_{60}^-$. 

The derived vibronic coupling parameters are listed in Table \ref{Table:V} and one of the fittings is shown in Fig. \ref{Fig:V} (see for the other fittings Supplemental Materials).
In the fitting, the $h_g\epsilon$ deformation is used instead of the $h_g\theta$ with $v_h = -(2/\sqrt{3}) \partial \epsilon_x/\partial q_{h\epsilon}|_{\bm{q}=0}$.
The stabilization energies (\ref{Eq:Umin}) in the $T_{1g}$ electronic states are $E_a = 24.0$ and $E_h = 65.6$ meV, which are by 30.7 \% and 32.5 \% larger than the stabilization energies of for the $a_{g}$ and $h_{g}$ modes in the $T_{1u}$ ground electronic states, respectively.

\subsection{Vibronic states}
\label{Sec:vibronic_results}
\begin{figure}[bt]
\includegraphics[width=8cm]{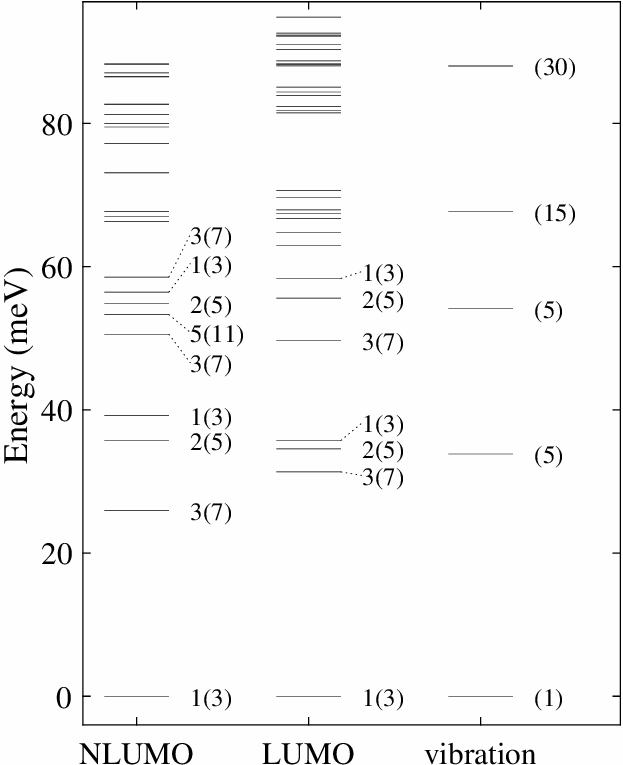}
\caption{Low-lying vibronic levels with respect to the first excited states, the ground states and zero-vibrational level of C$_{60}^{-}$. The numbers next to the energy levels are $J$ with the degeneracy shown in the parenthesis.}
\label{Fig:Vibronic_E_level}
\end{figure}

The vibronic Hamiltonian matrix was numerically diagonalized as described in Sec. \ref{Sec:vibronic}.
The calculated data are listed in Table \ref{Table:Energy_level} and the levels are shown in Fig. \ref{Fig:Vibronic_E_level}.
In the figure, the vibronic levels for the ground C$_{60}^-$ and the vibrational levels of neutral C$_{60}$ as well as the vibronic levels of the first excited C$_{60}^-$.
The ground vibronic and vibrational levels are used as the origin of the energy. 

One should note that the distributions of the vibronic states of C$_{60}$ anions differ much from that of the vibrational levels of neutral C$_{60}$.
The ground vibronic levels with vibronic angular momentum $J$=1 for the $T_{1u}$ ($t_{1u}^1$) and $T_{1g}$ ($t_{1g}^1$) electornic states are $-96.5$ and $-113.8$ meV, respectively. 
Previous study shows that for the ground electronic states, the contributions from the static and the dynamic JT effect to the ground energy are almost the same\cite{Liu2018a}. 
However, the ratio of the dynamical contribution to the static contribution is smaller in the $t_{1g}^1$ case than in the $t_{1u}^1$ case (Table \ref{Table:Energy_contrib}), 
which is consistent \cite{Auerbach1994, OBrien1996} with the stronger orbital vibronic couplings for the $t_{1g}$ NLUMO than for the $t_{1u}$ LUMO. 

The difference in the vibronic couplings in the $t_{1u}$ and the $t_{1g}$ levels appear in the excited vibronic levels too (Fig. \ref{Fig:Vibronic_E_level}).
The group of the first excited vibronic levels ($J = 3,2,1$) split more in the case of $t_{1g}^1$ than in $t_{1u}^1$, as expected from the stronger vibronic coupling in the former: the splitting of the former, 13.3 meV, is about two times larger than that of LUMO (4.4 meV).
Such splitting may be observed as fine structures in e.g. high-resolution absorption spectra of C$_{60}^-$.

\begin{table}[tb]
\small
\caption{
The vibronic energy levels with respect to the first excited states and the ground states of C$_{60}^{-}$ (meV). The data for are taken Ref. \cite{Liu2018b}, and the number in the parentheses indicate $J$.}
\label{Table:Energy_level}
\begin{tabular*}{0.48\textwidth}{@{\extracolsep{\fill}}cccc}
\hline
       $J$          &       & NLUMO & LUMO \\
                     \hline
\multirow{9}{*}{1} & 1 &$-113.815$& $-96.469$ \\
                     & 2 & $-74.589$& $-60.753$ \\
                     & 3 & $-57.370$& $-38.126$ \\
                     & 4 & $-47.520$& $-29.757$ \\
                     & 5 & $-40.736$& $-26.841$ \\
                     & 6 & $-31.168$& $-11.395$ \\
                     & 7 & $-25.493$&  $-8.099$ \\
                     & 8 &  -       &  $-5.411$ \\
                     & 9 &  -       &  $-4.123$ \\
                     \hline
\multirow{7}{*}{2} & 1 & $-78.132$& $-61.918$ \\
                     & 2 & $-58.955$& $-40.873$ \\
                     & 3 & $-46.127$& $-29.004$ \\
                     & 4 & $-31.122$& $-12.071$ \\
                     & 5 & $-26.748$&  $-8.264$ \\
                     & 6 & -        &  $-6.155$ \\
                     & 7 & -        &  $-4.265$ \\
                     \hline
\multirow{9}{*}{3} & 1 & $-87.854$& $-65.135$ \\
                     & 2 & $-63.252$& $-46.786$ \\
                     & 3 & $-55.282$& $-31.703$ \\
                     & 4 & $-40.633$& $-25.806$ \\
                     & 5 & $-33.815$& $-14.620$ \\
                     & 6 & $-32.593$& $-12.575$ \\
                     & 7 & $-25.581$&  $-8.417$ \\
                     & 8 & -        &  $-3.843$ \\
                     & 9 & -        &  $-1.607$ \\
                     \hline
\multirow{3}{*}{4} & 1 & $-46.772$& $-28.549$ \\
                     & 2 & $-34.314$& $-14.108$ \\
                     & 3 & $-27.326$& $-7.724$ \\
                     \hline
\multirow{3}{*}{5} & 1 & $-60.505$& $-33.554$ \\
                     & 2 & $-36.593$& $-14.985$ \\
                     & 3 & $-27.281$& $ -$ \\
                     \hline
\end{tabular*}
\end{table}

\begin{table}[tb]
\small
\caption{
Contributions to the ground vibronic energy (E$_{total}$) of NLUMO and LUMO of C$_{60}^{-}$. E$_{static}$, and E$_{dynamic}$ represent the static JT and dynamic JT stabilization energies. Ratio refers the ratio between E$_{static}$ and E$_{dynamic}$ (E$_{dynamic}$/E$_{static}$).}
\label{Table:Energy_contrib}
\begin{tabular*}{0.48\textwidth}{@{\extracolsep{\fill}}lllll}
\hline
  Orbital   &  E$_{total}$& E$_{static}$ & E$_{dynamic}$ & Ratio\\
\hline
NLUMO       & $-113.8$  &$-65.6$ &$-48.2$   & 0.74 \\
LUMO\cite{Liu2018a} &$-96.46$    &$-50.3$   &$-46.2$      & 0.92 \\
\hline
\end{tabular*}
\end{table}

\section{Discussion}
The dynamical JT effect in the $T_{1g}$ electronic states has been discussed in e.g. Refs. \onlinecite{Kondo1995, Tomita2005}. 
Indeed, as found in Sec. \ref{Sec:vibronic_results}, the stabilization by the dynamical JT effect is even stronger than in the ground electronic states (Table \ref{Table:Energy_contrib}), suggesting the importance of the vibronic dynamics in the excited electronic states.  

The presence of the dynamical JT effect in the excited states can be found in various spectroscopic data. 
For example, one should note that the signs of almost all the vibronic coupling parameters for the $h_g$ modes in the $T_{1g}$ state, and hence the JT deformations, are opposite to those for the $T_{1u}$ state.
The difference in the direction of the JT deformations in the $T_{1u}$ and the $T_{1g}$ states indicate that the relative displacements in these electronic states are large, and thus, the vibronic progression in the excitation spectra of ${}^2T_{1g} \leftarrow {}^2T_{1u}$ tends to be stronger than that in the photoelectron spectra of C$_{60}^-$ because the latter is only related to the vibronic coupling in the $T_{1u}$ state.  
Indeed, the the peaks in the absorption spectra of C$_{60}^-$ \cite{Kondo1995, Tomita2005} are stronger than the peaks in the photoelectron spectra of C$_{60}^-$ \cite{Wang2005, Huang2014}. 
Furthermore, as mentioned in Sec. \ref{Sec:vibronic_results}, the transition to the excited vibronic states may be seen as the fine structure of the spectra.  

By combining the present vibronic coupling parameters for the $t_{1g}$ NLUMO with those for the $t_{1u}$ LUMO (Table \ref{Table:V}), it is also possible to address diverse problems of C$_{60}$. 
For example, the $(T_{1u} \oplus T_{1g}) \otimes h_g$ product Jahn-Teller problem \cite{productJT} in C$_{60}^{2-}$ anion $t_{1u}^1t_{1g}^1$ electron configurations.
Since the signs of the orbital vibronic coupling parameters for the $t_{1u}$ and the $t_{1g}$ levels tend to be opposite to each other, the resulting vibornic coupling of the first excited C$_{60}^{2-}$ must be weak.
Besides, the combination of the present vibronic coupling constants and those for C$_{60}^+$ \cite{Huang2019c60plusarxiv} enables us to investigate the luminescence spectra \cite{Akimoto2002} involving $t_{1g}$ NLUMO.

\section{Conclusion}
In this work, the vibronic coupling parameters of the $T_{1g}$ electronic states of C$_{60}^-$ by using the next lowest unoccupied molecular orbital levels at B3LYP level. 
Based on the obtained parameters, the vibronic states were calculated by exactly diagonalizing the dynamical Jahn-Teller Hamiltonian. 
The results for the $t_{1g}^1$ configuration showed stronger dynamic JT stabilization than that for the $t_{1u}^-$ configuration by about 20 \%, indicating the importance of the JT effect in the excited states. 
The presence of the JT dynamics appears in spectroscopic data. 
Due to the difference in the direction of the JT deformations in the $T_{1u}$ and the $T_{1g}$ states, stronger vibronic progression is seen in the absorption spectra than in the photoelectron spectra of C$_{60}^-$.

\begin{acknowledgments}
The authors thank Naoya Iwahara and Liviu Chibotaru for fruitful discussions. They also gratefully acknowledge funding by the China Scholarship Council (CSC).
\end{acknowledgments}

\bibliography{aiptemplate}

\end{document}


\title{Supplement Material: Dynamical Jahn-Teller effect in the first excited C$_{60}^-$}
\author{Zhishuo Huang}
\email{zhishuohuang@gmail.com}
\affiliation{Theory of Nanomaterials Group, KU Leuven, Celestijnenlaan 200F, B-3001 Leuven, Belgium}
\author{Dan Liu}
\email{iamdliu@nwpu.edu.cn}
\affiliation{Shaanxi Institute of Flexible Electronics, Northwestern Polytechnical University, 127 West Youyi Road, Xi'an, 710072, Shaanxi, China}
\affiliation{Theory of Nanomaterials Group, KU Leuven, Celestijnenlaan 200F, B-3001 Leuven, Belgium}
\date{\today}

\maketitle
\onecolumngrid
  Supplemental Materials contain the JT splitting of the NLUMO levels with respect to $q_{h\epsilon}$ deformation.

  \section{JT splitting of the NLUMO levels}
\begin{figure}[bt]
\includegraphics[width=10cm]{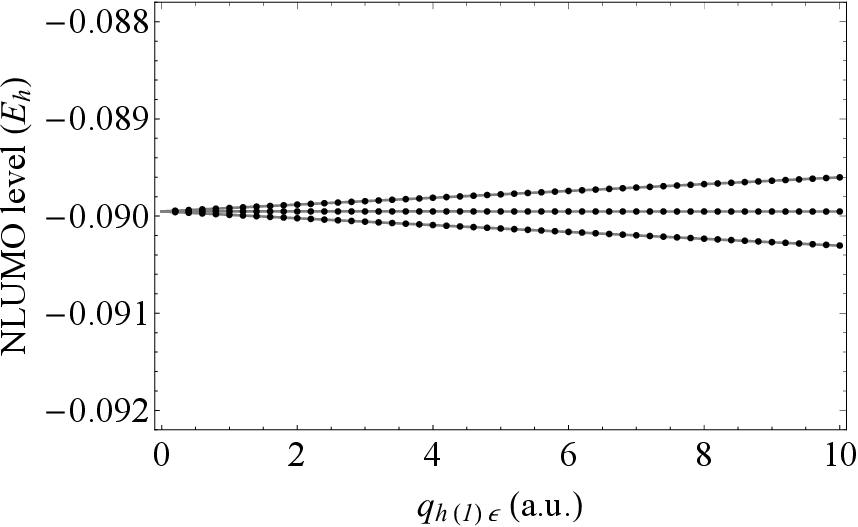}
\caption{
The JT splitting of the NLUMO levels with respect to $q_{h_g(1)\epsilon}$ deformation (in atomic unit).
The black points and gray lines indicate the DFT values and model energy, respectively.
}
\label{Fig:Vh1}
\end{figure}

\begin{figure}[bt]
\includegraphics[width=10cm]{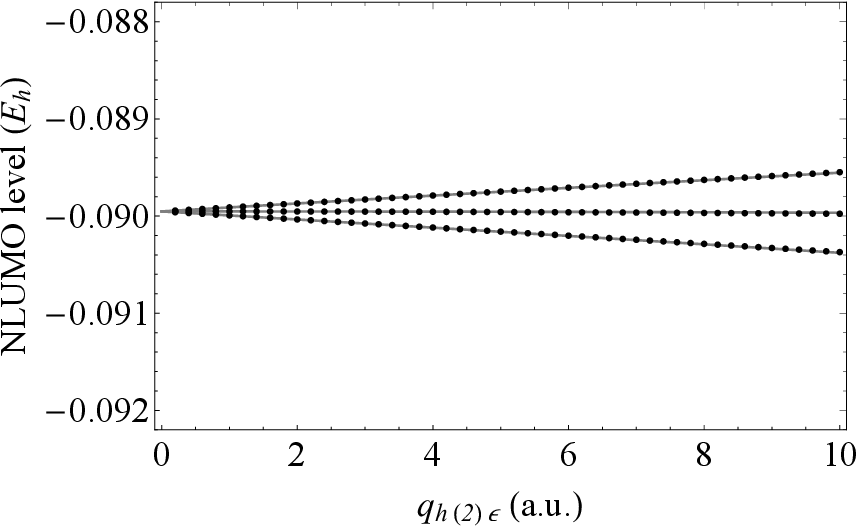}
\caption{
The JT splitting of the NLUMO levels with respect to $q_{h_g(2)\epsilon}$ deformation (in atomic unit).
The black points and gray lines indicate the DFT values and model energy, respectively.
}
\label{Fig:Vh2}
\end{figure}

\begin{figure}[bt]
\includegraphics[width=10cm]{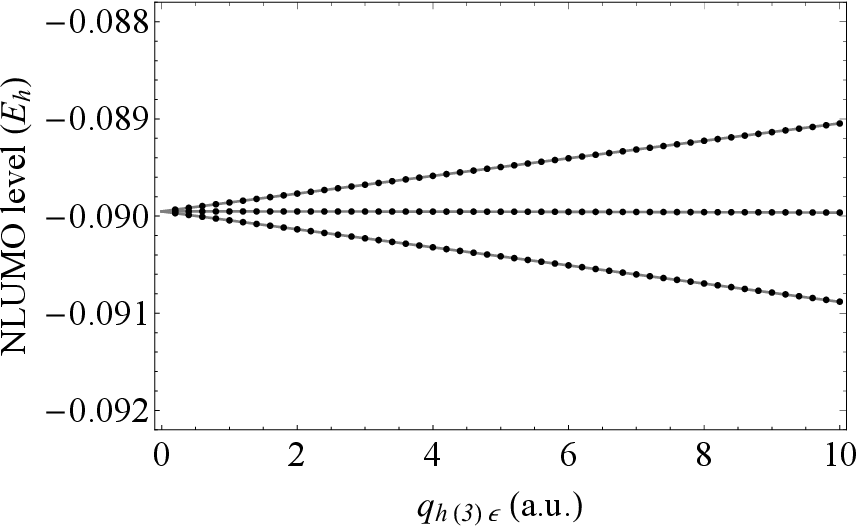}
\caption{
The JT splitting of the NLUMO levels with respect to $q_{h_g(3)\epsilon}$ deformation (in atomic unit).
The black points and gray lines indicate the DFT values and model energy, respectively.
}
\label{Fig:Vh3}
\end{figure}

\begin{figure}[bt]
\includegraphics[width=10cm]{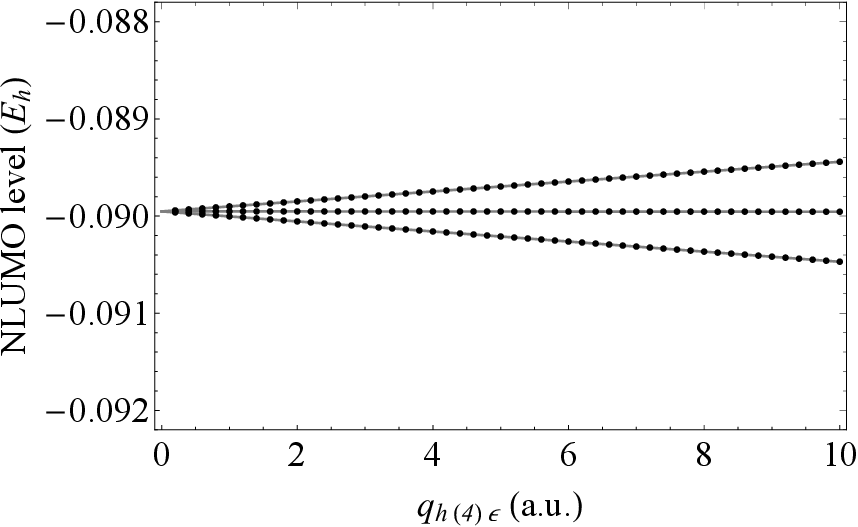}
\caption{
The JT splitting of the NLUMO levels with respect to $q_{h_g(4)\epsilon}$ deformation (in atomic unit).
The black points and gray lines indicate the DFT values and model energy, respectively.
}
\label{Fig:Vh4}
\end{figure}

\begin{figure}[bt]
\includegraphics[width=10cm]{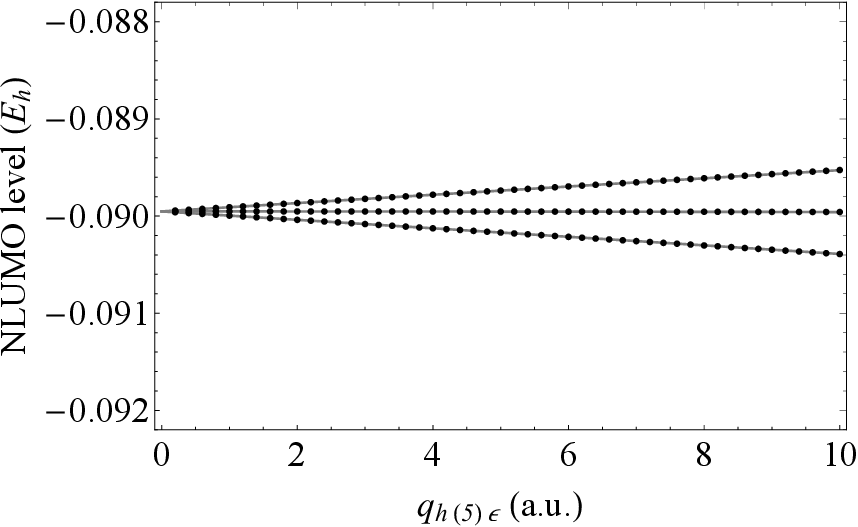}
\caption{
The JT splitting of the NLUMO levels with respect to $q_{h_g(5)\epsilon}$ deformation (in atomic unit).
The black points and gray lines indicate the DFT values and model energy, respectively.
}
\label{Fig:Vh5}
\end{figure}

\begin{figure}[bt]
\includegraphics[width=10cm]{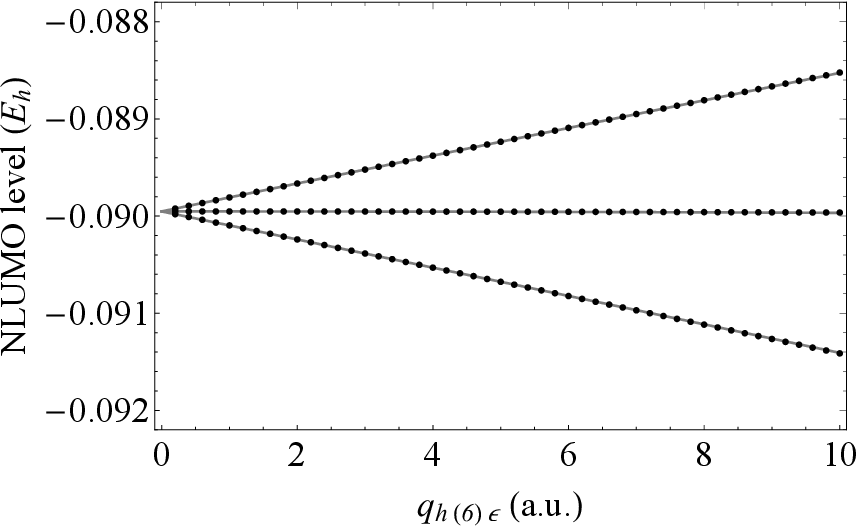}
\caption{
The JT splitting of the NLUMO levels with respect to $q_{h_g(6)\epsilon}$ deformation (in atomic unit).
The black points and gray lines indicate the DFT values and model energy, respectively.
}
\label{Fig:Vh6}
\end{figure}

\begin{figure}[bt]
\includegraphics[width=10cm]{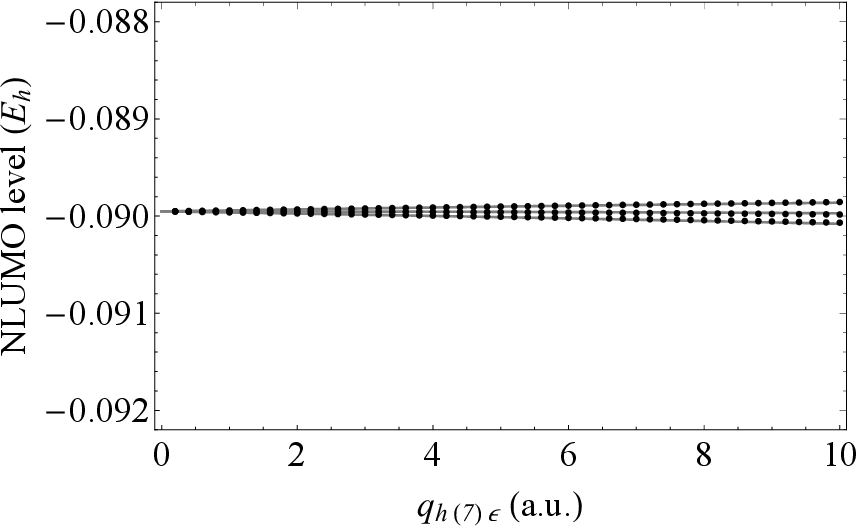}
\caption{
The JT splitting of the NLUMO levels with respect to $q_{h_g(7)\epsilon}$ deformation (in atomic unit).
The black points and gray lines indicate the DFT values and model energy, respectively.
}
\label{Fig:Vh7}
\end{figure}

\begin{figure}[bt]
\includegraphics[width=10cm]{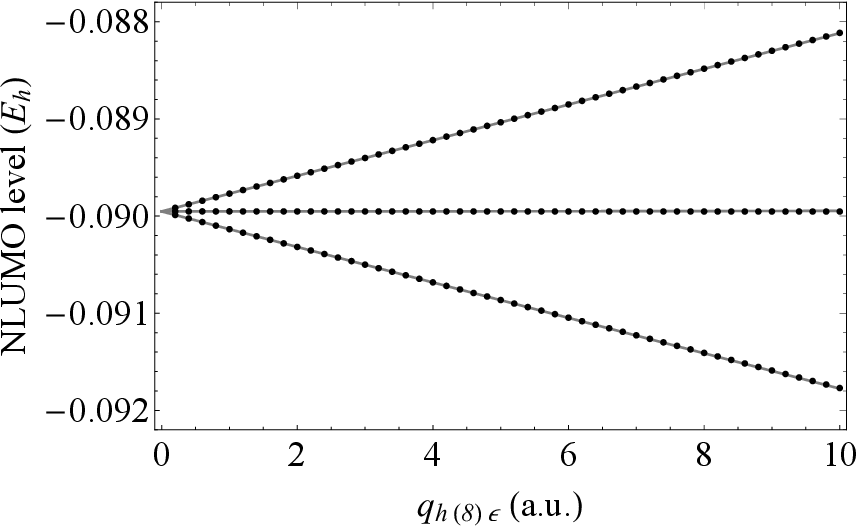}
\caption{
The JT splitting of the NLUMO levels with respect to $q_{h_g(8)\epsilon}$ deformation (in atomic unit).
The black points and gray lines indicate the DFT values and model energy, respectively.
}
\label{Fig:Vh8}
\end{figure}

    There are eight $q_{h\epsilon}$ deformation, which are distinguished by the subindex, as $q_{h(1)\epsilon}$ corresponding to the first $q_{h\epsilon}$ deformation. The splitting patten of the NLUMO levels with respect to $q_{h(i)\epsilon}$ (i=1,2...8) deformation are shown in Fig. \ref{Fig:Vh1},  \ref{Fig:Vh2}, \ref{Fig:Vh3}, \ref{Fig:Vh4}, \ref{Fig:Vh5}, \ref{Fig:Vh6}, \ref{Fig:Vh7}, and \ref{Fig:Vh8}, respectively.


